\documentclass[11pt]{article} 
\usepackage{hyperref}

\usepackage{geometry}
\usepackage{amsthm}

\usepackage[round]{natbib}
\usepackage{amsbsy}
\usepackage{amsmath}    
\usepackage{amssymb}
\usepackage{booktabs}
\usepackage[figuresright]{rotating}

\providecommand{\keywords}[1]{\textbf{\textit{Keywords---}} #1}

\begin{document}

\title{Nonparametric Stein-type Shrinkage Covariance Matrix Estimators in High-Dimensional Settings}
\author{
Anestis Touloumis\\ 
Cancer Research UK Cambridge Institute\\
 University of Cambridge\\
 Cambridge CB2 0RE, U.K.\\
\texttt{Anestis.Touloumis@cruk.cam.ac.uk} 
}
\date{}
\maketitle

\begin{abstract}
Estimating a covariance matrix is an important task in applications where the number of variables is larger than the number of observations. In the literature, shrinkage approaches for estimating a high-dimensional covariance matrix are employed to circumvent the limitations of the sample covariance matrix. A new family of nonparametric Stein-type shrinkage covariance estimators is proposed whose members are written as a convex linear combination of the sample covariance matrix and of a predefined invertible target matrix. Under the Frobenius norm criterion, the optimal shrinkage intensity that defines the best convex linear combination depends on the unobserved covariance matrix and it must be estimated from the data. A simple but effective estimation process that produces nonparametric and consistent estimators of the optimal shrinkage intensity for three popular target matrices is introduced. In simulations, the proposed Stein-type shrinkage covariance matrix estimator based on a scaled identity matrix appeared to be up to $80\%$ more efficient than existing ones in extreme high-dimensional settings. A colon cancer dataset was analyzed to demonstrate the utility of the proposed estimators. A rule of thumb for adhoc selection among the three commonly used target matrices is recommended.
\end{abstract}


\section{Introduction}\label{Sec00}
The problem of estimating large covariance matrices arises frequently in modern applications, such as in genomics, cancer research, clinical trials, signal processing, financial mathematics, pattern recognition and computational convex geometry. Formally, the goal is to estimate the covariance matrix $\boldsymbol \Sigma$ based on a sample of $N$ independent and identically distributed (i.i.d) $p$-variate random vectors $\mathbf X_{1},\ldots,\mathbf X_{N}$ with mean vector $\boldsymbol \mu$ in the ``small $N$, large $p$'' paradigm, that is when $N$ is a lot smaller compared to $p$. It is a well-known fact that the sample covariance matrix 
$$\mathbf S=\frac{1}{N-1} \sum_{i=1}^N (\mathbf X_i-\mathbf{\bar{X}})(\mathbf X_i-\mathbf{\bar{X}})^T,$$
where $\mathbf{\bar{X}}=\sum_{i=1}^N \mathbf X_i/N$ is the sample mean vector, is not performing satisfactory in high-dimensional settings. For example, $\mathbf S$ is singular even when $\boldsymbol \Sigma$ is a strictly positive definite matrix. Recent research in estimating high-dimensional covariance matrices includes banding, tapering, penalization and shrinkage methods. We focus on the Steinian shrinkage method \citep{Stein1956} as adopted by \cite{Ledoit2004} because it leads to covariance matrix estimators that are: (i) non-singular (ii) well-conditioned, (iii) invariant to permutations of the order of the $p$ variables, (iv) consistent to departures from a multivariate normal model, (v) not necessarily sparse, (vi) expressed in closed form and (vii) computationally cheap regardless of $p$. 

\indent \cite{Ledoit2004} proposed a Stein-type covariance matrix estimator for $\boldsymbol \Sigma$ based on 
\begin{equation}
\mathbf S^{\star} = (1-\lambda) \mathbf S + \lambda \nu \mathbf I_{p},
\label{bonafide}
\end{equation}
where $\mathbf I_{p}$ is the $p \times p$ identity matrix, and where $\lambda$ and $\nu$ minimize the risk function $\mathrm{E}\left[||\mathbf S^{\star} -\boldsymbol \Sigma||_F^2\right]$, that is
$$\lambda=\frac{\mathrm{E}\left[||\mathbf S -\boldsymbol \Sigma||^2_F\right]}{\mathrm{E}\left[||\mathbf S- \nu \mathbf I_{p}||^2_F\right]}$$
and
$$\nu=\frac{\mathrm{tr}(\boldsymbol \Sigma)}{p}.$$ 
The optimal shrinkage intensity parameter $\lambda$ in~(\ref{bonafide}) suggests how much we must shrink the eigenvalues of the sample covariance matrix $\mathbf S$ towards the eigenvalues of the target matrix $\nu \mathbf I_{p}$. For example, $\lambda=0$ implies no contribution of $\nu \mathbf I_{p}$ to $\mathbf S^{\star}$, while $\lambda=1$ implies no contribution of $\mathbf S$ to $\mathbf S^{\star}$. Intermediate values for $\lambda$ reveal the simultaneous contribution of $\mathbf S$ and $\nu \mathbf I_{p}$ to $\mathbf S^{\star}$. Despite the attractive interpretation, $\mathbf S^{\star}$ is not a covariance matrix estimator because $\nu$ and $\lambda$ depend on the unobservable covariance matrix $\boldsymbol \Sigma$. For this reason, \cite{Ledoit2004} proposed to plug-in nonparametric $N$-consistent estimators for $\nu$ and $\lambda$ in~(\ref{bonafide}) and use the resulting matrix as a shrinkage covariance matrix estimator for $\boldsymbol \Sigma$. Although $\nu$ seems to be adequately estimated by $\hat{\nu}=\mathrm{tr}(\mathbf S)/p$, we noticed via simulations that the estimator of $\lambda$ proposed by \cite{Ledoit2004} was biased in extreme high-dimensional settings and when $\boldsymbol \Sigma=\mathbf I_p$. This is counter-intuitive because $\lambda=1$ and the plug-in estimator of $\mathbf S^{\star}$ is expected to be as close as possible to the target matrix $\nu \mathbf I_{p}$. In addition, this observation underlines the importance of choosing a target matrix that approximates well the true underlying dependence structure. To this direction, \cite{Fisher2011} proposed Stein-type shrinkage covariance matrix estimators for alternative target matrices. However, they are no longer nonparametric as their construction was based on a multivariate normal model assumption. 
  
\indent Motivated by the above, we improve estimation of the optimal shrinkage intensity by providing a consistent estimator of $\lambda$ in high-dimensional settings. To construct the estimator of $\lambda$ we follow three simple steps: (i) expand the expectations in the numerator and denominator of $\lambda$ assuming a multivariate normal model, (ii) prove that this ratio, say $\lambda^{\star}$, is asymptotically equivalent to $\lambda$, and (iii) replace each unknown parameter in $\lambda^{\star}$ with unbiased and consistent estimators constructed using $U$-statistics. The last step is essential in our proposal so as to ensure consistent and nonparametric estimation of $\lambda$. Further, we relax the normality assumption in \cite{Fisher2011} for target matrices other than $\nu \mathbf I_{p}$ in (\ref{bonafide}) and we illustrate how to estimate consistently the corresponding optimal shrinkage intensities in high-dimensional settings. In other words, we propose a new nonparametric family of Stein-type shrinkage estimators suitable for high-dimensional settings that preserve the attractive properties mentioned in the first paragraph and can accommodate arbitrary target matrices.

\indent The rest of this paper is organized as follows. In Section \ref{Sec01}, we present the working framework that allows us to manage the high-dimensional setting. Section \ref{Sec02} contains the main results where we derive consistent and nonparametric estimators for the optimal shrinkage intensity of three different target matrices. We evaluate the performance of the proposed covariance matrix estimators via simulations in Section \ref{Sec03}. In Section \ref{Sec04}, we illustrate the use of the proposed estimators in a colon cancer study and we recommend a rule of thumb for selecting the target matrix. In Section \ref{Sec05}, we summarize our findings and discuss future research. The technical details can be found in the appendix. Throughout the paper, we use $||\mathbf A||^2_{F}=\mathrm{tr}(\mathbf A^{T}\mathbf A)/p$ to denote the scaled Frobenius norm of $\mathbf A$, $\mathrm{tr}(\mathbf A)$ to denote the trace of the matrix $\mathbf A$, $\mathbf D_{\mathbf A}$ to denote the diagonal matrix with elements the diagonal elements of $\mathbf A$, and $\mathbf A \circ \mathbf B$ to denote the Hadamard product of the matrices $\mathbf A$ and $\mathbf B$, i.e., the matrix whose $(a,b)$-th element is the product of the corresponding elements of $\mathbf A$ and $\mathbf B$. In the above, it is implicit that $\mathbf A$ and $\mathbf B$ are $p \times p$ matrices.

\section{Framework for High-Dimensional Settings} \label{Sec01}
Let $\mathbf X_{1},\ldots,\mathbf X_{N}$ be a sample of i.i.d. $p$-variate random vectors from the nonparametric model
\begin{equation}
\mathbf X_{i} = {\boldsymbol \Sigma}^{1/2} \mathbf Z_{i} + \boldsymbol \mu,
\label{Nonparametricmodel}
\end{equation}
where $\boldsymbol \mu=\mathrm{E}[\mathbf X_i]$ is the $p$-variate mean vector, $\boldsymbol \Sigma=\mathrm{cov}[\mathbf X_{i}]=\boldsymbol \Sigma^{1/2} \boldsymbol \Sigma^{1/2}$ is the $p \times p$ covariance matrix, and $\mathbf Z_{1},\ldots,\mathbf Z_{N}$ is a collection of i.i.d. $p$-variate random vectors. Instead of distributional assumptions, moments restrictions are imposed on the random variables in $\mathbf Z_i$. In particular, let $Z_{ia}$ be the $a$-th random variable in $\mathbf Z_i$ and suppose that $\mathrm{E}[Z_{ia}]=0$, $\mathrm{E}[Z^{2}_{ia}]=1$, $\mathrm{E}[Z^4_{ia}]=3+B$ with $-2 \leq B < \infty$ and for any nonnegative integers $l_1,\ldots,l_4$ such that $\sum_{\nu=1}^4 l_{\nu} \leq 4$ 
\begin{equation}
\mathrm{E}[Z^{l_1}_{ia_1} Z^{l_2}_{ia_2} Z^{l_3}_{ia_3} Z^{l_4}_{ia_4}]=\mathrm{E}[Z^{l_1}_{ia_1}]\mathrm{E}[Z^{l_2}_{ia_2}] \mathrm{E}[Z^{l_2}_{ia_3}] \mathrm{E}[Z^{l_4}_{ia_4}],
\label{qindependence}
\end{equation}
where the indexes $a_1,\ldots,a_4$ are distinct. The nonparametric model~(\ref{Nonparametricmodel}) includes the $p$-variate normal distribution $\mathrm{N}_p(\boldsymbol \mu, \boldsymbol \Sigma)$ as a special case obtained if $Z_{ia}$ are i.i.d. $\mathrm{N}(0,1)$ random variables. Since $B=0$ under a multivariate normal model, $B$ can be interpreted as a measure of departure of the fourth moment of $Z_{ia}$ to that of a $\mathrm{N}(0,1)$ random variable. The assumption of common fourth moments is made for notational ease and the results of this paper remain valid even if $E[Z^4_{ia}]=3+B_a$ for finite $B_a$ ($a=1,\ldots,p$). Model~(\ref{Nonparametricmodel}) assumes that $\mathbf X_{i}$ is a linear combination of $\mathbf Z_i$, a random vector that contains standardized white noise variables. Unlike the usual definition of white noise, model~(\ref{Nonparametricmodel}) makes no distributional assumptions for $\mathbf Z_i$ and it allows dependence patterns given that the pseudo-independence condition~(\ref{qindependence}) holds. Therefore, the working framework can cover situations in which the white noise mechanism does not produce independent and/or identically distributed random variables. 

\indent To handle the high-dimensional setting, we restrict the dimension of $\boldsymbol \Sigma$ by assuming that
\begin{equation}
\text{as } N \rightarrow \infty, p=p(N) \rightarrow \infty, \frac{\mathrm{tr}(\boldsymbol \Sigma^4)}{\mathrm{tr}^2(\boldsymbol \Sigma^2)}\rightarrow t_1 \text{ and } \frac{\mathrm{tr}(\boldsymbol \Sigma^2)}{\mathrm{tr}^2(\boldsymbol \Sigma)}\rightarrow t_2, 
\label{CovMatAss}
\end{equation}
with $0\leq t_2 \leq t_1 \leq 1$. This flexible assumption does not specify the limiting behavior of $p$ with respect to $N$, thus including the case where $p/N$ is bounded \citep{Ledoit2004}. At the same time, it does not seriously restrict the class of covariance matrices for $\boldsymbol \Sigma$ that satisfy assumption~(\ref{CovMatAss}). For example, members of this class are covariance matrices whose eigenvalues are bounded away from $0$ and $\infty$ \citep{Chen2010a}, banded first-order auto-regressive covariance matrices such that $\boldsymbol \Sigma=\{\sigma_a\sigma_b\rho^{|a-b|}I(|a-b|\leq k)\}_{1 \leq a,b \leq p}$ with $-1\leq \rho \leq 1$, $\sigma_a$ and $\sigma_b$ bounded positive constants and $1 \leq k \leq p$ \citep{Chen2010a}, covariance matrices that have a few divergent eigenvalues as long as they diverge slowly \citep{Chen2010}, and covariance matrices with a compound symmetry correlation pattern, that is $\boldsymbol \Sigma=\{\sigma_a\sigma_b\rho\}_{1 \leq a,b \leq p}$. Model~(\ref{Nonparametricmodel}) and assumption (\ref{CovMatAss}) constitute an attractive working framework to handle the `small $N$, large $p$' paradigm. A similar but stricter working framework was considered by \cite{Chen2010a} in the context of hypothesis testing for $\boldsymbol \Sigma$. By contrast, we avoid making assumptions about moments of fifth or higher order and we allow more options for $\boldsymbol \Sigma$. 

\section{Main Results}\label{Sec02}
Under the nonparametric model~(\ref{Nonparametricmodel}), the expectations in the numerator and denominator of the optimal shrinkage intensity $\lambda$ in~(\ref{bonafide}) can be explicitly calculated to obtain that
\begin{equation*}
\lambda=\frac{\mathrm{E}\left[||\mathbf S -\boldsymbol \Sigma||^2_F\right]}{\mathrm{E}\left[||\mathbf S- \nu \mathbf I||^2_F\right]} =\frac{\mathrm{tr}(\boldsymbol \Sigma^2)+\mathrm{tr}^2(\boldsymbol \Sigma)+B \mathrm{tr}(\mathbf D^2_{\boldsymbol \Sigma})}{N\mathrm{tr}(\boldsymbol \Sigma^2)+\frac{p-N+1}{p}\mathrm{tr}^2(\boldsymbol \Sigma)+B \mathrm{tr}(\mathbf D^2_{\boldsymbol \Sigma})}.
\label{optimal.equal}
\end{equation*}
Since $\mathrm{tr}(\mathbf D^2_{\mathbf A})=\mathrm{tr}(\mathbf A \circ \mathbf A) \leq \mathrm{tr}(\mathbf A^2) \leq \mathrm{tr}^2(\mathbf A)$ for any positive definite matrix $\mathbf A$, the contribution of $B \mathrm{tr}(\mathbf D^2_{\boldsymbol \Sigma})$ to $\lambda$ under assumption~(\ref{CovMatAss}) is negligible when compared to that of $\mathrm{tr}(\boldsymbol \Sigma^2)$ or $\mathrm{tr}^2(\boldsymbol \Sigma)$. Ignore $B\mathrm{tr}(\mathbf D^2_{\boldsymbol \Sigma})$ in $\lambda$ and define
\begin{equation*}
\lambda^{\star}=\frac{\mathrm{tr}(\boldsymbol \Sigma^2)+\mathrm{tr}^2(\boldsymbol \Sigma)}{N\mathrm{tr}(\boldsymbol \Sigma^2)+\frac{p-N+1}{p}\mathrm{tr}^2(\boldsymbol \Sigma)}.
\end{equation*}
This is the optimal shrinkage intensity of the multivariate normal model or, more generally, $\lambda=\lambda^{\star}$ when $B=0$ in the nonparametric model~(\ref{Nonparametricmodel}). Under assumption~(\ref{CovMatAss}), it follows that 
\begin{align*}
|\lambda-\lambda^{\star}|&=\frac{(N-1) \left(\mathrm{tr}(\boldsymbol \Sigma^2)-\frac{1}{p}\mathrm{tr}^2(\boldsymbol \Sigma)\right)}{N \left(\mathrm{tr}(\boldsymbol \Sigma^2)-\frac{1}{p}\mathrm{tr}^2(\boldsymbol \Sigma)\right)+\frac{p+1}{p}\mathrm{tr}^2(\boldsymbol \Sigma)}\\
                         &\times\frac{|B| \mathrm{tr}(\mathbf D^2_{\boldsymbol \Sigma})}{N \left(\mathrm{tr}(\boldsymbol \Sigma^2)-\frac{1}{p}\mathrm{tr}^2(\boldsymbol \Sigma)\right)+\frac{p+1}{p}\mathrm{tr}^2(\boldsymbol \Sigma)+B\mathrm{tr}(\mathbf D^2_{\boldsymbol \Sigma})}\\
                         &\leq \frac{|B| \frac{\mathrm{tr}(\mathbf D^2_{\boldsymbol \Sigma})}{\mathrm{tr}^2(\boldsymbol \Sigma)}}{N \left(\frac{\mathrm{tr}(\boldsymbol \Sigma^2)}{\mathrm{tr}^2(\boldsymbol \Sigma)}-\frac{1}{p}\right)+\frac{p+1}{p}+B\frac{\mathrm{tr}(\mathbf D^2_{\boldsymbol \Sigma})}{\mathrm{tr}^2(\boldsymbol \Sigma)}} \rightarrow 0,
\end{align*}
where $|k|$ denotes the absolute value of the real number $k$. Therefore, the optimal shrinkage intensity obtained under normality is asymptotically equivalent to the optimal shrinkage intensity with respect to model~(\ref{Nonparametricmodel}) as long as the trace ratios restrictions in~(\ref{CovMatAss}) hold. This means that it suffices to construct a nonparametric and consistent estimator of $\lambda^{\ast}$ to estimate $\lambda$. To accomplish this, replace the unknown parameters $\mathrm{tr}(\boldsymbol \Sigma)$ and $\mathrm{tr}(\boldsymbol \Sigma^2)$ in $\lambda^{\star}$ with the unbiased estimators 
\begin{equation*}
Y_{1N}=U_{1N}-U_{4N}=\frac{1}{N}\sum_{i=1}^N \mathbf X^{T}_{i} \mathbf X_{i} -\frac{1}{P^N_2}\sum_{i \neq j}^{\ast} \mathbf X^{T}_{j}\mathbf X_{i}
\end{equation*}
and 
\begin{align*}
Y_{2N}&=U_{2N}-2U_{5N}+U_{6N}\nonumber\\
      &=\frac{1}{P^N_2}\sum_{i \neq j}^{\ast} (\mathbf X^{T}_{i}\mathbf X_{j})^2-2\frac{1}{P^N_3}\sum_{i \neq j \neq k}^{\ast} \mathbf X^{T}_{i}\mathbf X_{j}\mathbf X^{T}_{i}\mathbf X_{k}+\frac{1}{P^N_4} \sum_{i\neq j \neq k \neq l}^{\ast} \mathbf X_{i}\mathbf X^{T}_{j}\mathbf X_{k}\mathbf X^{T}_{l}
\end{align*}
respectively, where $P^s_t=s!/(s-t)!$ and $\sum^{*}$ denotes summation over mutually distinct indices. Note that $U_{1N}$ and $U_{2N}$ are the unbiased estimators of $\mathrm{tr}(\boldsymbol \Sigma)$ and $\mathrm{tr}(\boldsymbol \Sigma^2)$ respectively, when the data are centered around the mean vector (i.e., $\boldsymbol \mu=\mathbf 0$), while the remaining terms ($U_4$,$U_5$ and $U_6$) are $U$-statistics of second, third and fourth order that ensure the unbiasedness of $Y_{1N}$ and $Y_{2N}$ when $\boldsymbol \mu \neq \mathbf 0$. In~\ref{Appendix2}, we argue that $Y_{1N}$ and $Y_{2N}$ are ratio-consistent estimators to $\mathrm{tr}(\boldsymbol \Sigma)$ and $\mathrm{tr}(\boldsymbol \Sigma^2)$ respectively. Here, it should be noted a statistic $\hat{\theta}$ is called a ratio-consistent estimator to $\theta$ if $\hat{\theta}/\theta$ converges in probability to one. Therefore, it follows from the continuous mapping theorem that
$$\hat{\lambda}=\frac{Y_{2N}+Y^2_{1N}}{NY_{2N}+\frac{p-N+1}{p}Y^2_{1N}}$$
is a consistent estimator of $\lambda$. The proposed Stein-type shrinkage estimator for $\boldsymbol \Sigma$,
$$\hat{\mathbf S}^{\star} = (1-\hat{\lambda}) \mathbf S + \hat{\lambda} \hat{\nu} \mathbf I_{p},$$
is obtained by plugging-in $\hat{\nu}=Y_{1N}/p$ and $\hat{\lambda}$ in (\ref{bonafide}). 

\subsection{Alternative target matrices}\label{AltMat}
Next, we consider target matrices other than $\nu \mathbf I_p$ in (\ref{bonafide}). This extension is motivated by situations where $\lambda \rightarrow 1$ and $\nu \mathbf I_p$ is not a good approximation of $\boldsymbol \Sigma$. In this case, $\hat{\mathbf S}^{\star}$ remains a well-defined and non-singular covariance matrix estimator but it fails to reflect the underlying dependence structure.

\indent Let $\mathbf T$ be a well-conditioned and non-singular target matrix, and define the matrix
\begin{equation}
\mathbf S_T^{\star} = (1-\lambda_{T}) \mathbf S + \lambda_{T} \mathbf T.
\label{bfmatrix}
\end{equation}
Simple algebraic manipulation shows that the optimal shrinkage intensity
\begin{equation}
\lambda_{T}=\frac{\mathrm{E}\left[||\mathbf S -\boldsymbol \Sigma||^2_F\right]+\mathrm{E}\left[\mathrm {tr}\{(\mathbf S -\boldsymbol \Sigma)(\mathbf \Sigma- \mathbf T)\}\right]/p}{\mathrm{E}\left[||\mathbf S- \mathbf T||^2_F\right]}
\label{lambdageneric}
\end{equation}
minimizes the expected risk function $\mathrm{E}[||\mathbf S_{T}^{\ast} -\boldsymbol \Sigma||^2_F]$. A closed form solution for $\lambda_{T}$ can be derived by calculating the expectations in~(\ref{lambdageneric}) with respect to model~(\ref{Nonparametricmodel}). The key idea of our proposal is to simplify the estimation process for $\lambda_T$ by identifying terms in the numerator and denominator of $\lambda_{T}$ that can be safely ignored under assumption~(\ref{CovMatAss}). One approach is to examine whether the optimal shrinkage intensity under the multivariate normal model assumption is asymptotically equivalent to $\lambda_T$. Whenever this is the case, we can set $B=0$ in $\lambda_{T}$ and replace the remaining parameters with unbiased and ratio-consistent estimators to obtain $\hat{\lambda}_{T}$. The proposed Stein-type shrinkage covariance matrix estimator is 
$$\hat{\mathbf S}_T^{\star} = (1-\hat{\lambda}_{T}) \mathbf S + \hat{\lambda}_{T} \mathbf T.$$
Note that if we set $\mathbf T=\nu \mathbf I_p$ then $\lambda_T=\lambda$ and thus $\mathbf S_T^{\star}=\mathbf S^{\star}$. We provide the estimator of $\lambda_{T}$ for two target matrices: i) the identity matrix $\mathbf T=\mathbf I_p$, and ii) the diagonal matrix $\mathbf T=\mathbf D_{\mathbf S}$ whose diagonal elements are the sample variances. To guarantee the consistency of $\lambda_{T}$, we suppose that $t_1=t_2=0$ in~(\ref{CovMatAss}). This assumption does not heavily affects the class of $\boldsymbol \Sigma$ under consideration. In fact, all the dependence patterns mentioned in Section~\ref{Sec01} satisfy this stronger version of assumption~(\ref{CovMatAss}) except the compound symmetry correlation matrix. We believe that this is a small price to pay if we are willing to increase the range of the target matrices in~(\ref{bonafide}). 

\indent First, when $\mathbf T=\mathbf I_p$ in~(\ref{bfmatrix}) the optimal shrinkage intensity in~(\ref{lambdageneric}) becomes
\begin{equation*}
\lambda_{I}=\frac{\mathrm{tr}(\boldsymbol \Sigma^2)+\mathrm{tr}^2(\boldsymbol \Sigma)+B \mathrm{tr}(\mathbf D^2_{\boldsymbol \Sigma})}{\mathrm{tr}(\boldsymbol \Sigma^2)+\mathrm{tr}^2(\boldsymbol \Sigma)+(N-1)\mathrm{tr}\left[(\boldsymbol \Sigma-\mathbf I_p)^2\right]+B \mathrm{tr}(\mathbf D^2_{\boldsymbol \Sigma})}.
\end{equation*}
As before, we can ignore the terms $B \mathrm{tr}(\mathbf D^2_{\boldsymbol \Sigma})$ in $\lambda_{I}$ and prove that 
\begin{equation*}
\hat{\lambda}_{I}=\frac{Y_{2N}+Y^2_{1N}}{NY_{2N}+Y^2_{1N}-(N-1)(2Y_{1N}-p)}
\end{equation*}
is a consistent estimator of $\lambda_{I}$.

\indent When $\mathbf T=\mathbf D_{\mathbf S}$, we can use the results in \ref{Appendix1} and prove that the optimal shrinkage intensity in~(\ref{lambdageneric}) is
\begin{align*}
\lambda_{D} &=\frac{\mathrm{E}\left[||\mathbf S -\boldsymbol \Sigma||^2_F\right]+\mathrm{E}\left[\mathrm {tr}\{(\mathbf S -\boldsymbol \Sigma)(\mathbf \Sigma- \mathbf D_{\mathbf S})\}\right]/p}{\mathrm{E}\left[||\mathbf S- \mathbf D_{\mathbf S}||^2_F\right]}\\
            &=\frac{\left(\mathrm{E}[\mathrm{tr}(\mathbf S^2)]-\mathrm{tr}(\mathbf \Sigma^2)\right)+\left(\mathrm{E}[\mathrm{tr}(\mathbf \Sigma \mathbf D_{\mathbf S})]-\mathrm{E}[\mathrm{tr}(\mathbf S\mathbf D_{\mathbf S})]\right)}{\mathrm{E}[\mathrm{tr}(\mathbf S^2)]-\mathrm{E}[\mathrm{tr}(\mathbf D^2_{\mathbf S})]}\\
            &=\frac{\mathrm{tr}(\boldsymbol \Sigma^2)+\mathrm{tr}^2(\boldsymbol \Sigma)-2 \mathrm{tr}(\mathbf D^2_{\mathbf \Sigma})+B\left[\mathrm{tr}(\mathbf D^2_{\mathbf \Sigma})-\sum_{a=1}^p\sum_{b=1}^p \left(\Sigma^{1/2}_{ab}\right)^4\right]}{N\mathrm{tr}(\boldsymbol \Sigma^2)+\mathrm{tr}^2(\boldsymbol \Sigma)-(N+1) \mathrm{tr}(\mathbf D^2_{\mathbf \Sigma})+B\left[\mathrm{tr}(\mathbf D^2_{\mathbf \Sigma}) -\sum_{a=1}^p\sum_{b=1}^p \left(\Sigma^{1/2}_{ab}\right)^4\right]},
\end{align*}
where $\Sigma^{1/2}_{ab}$ denotes the $(a,b)$-th element of $\mathbf \Sigma^{1/2}$. It can be shown that $B\left[\mathrm{tr}(\mathbf D^2_{\mathbf \Sigma})-\sum_{a=1}^p\sum_{b=1}^p \left(\Sigma^{1/2}_{ab}\right)^4\right]$ has negligible contribution to $\lambda_{D}$ and that 
\begin{align*}
Y_{3N}&=U_{3N}-2U_{7N}+U_{8N} \nonumber \\
      &=\frac{1}{P^N_2}\sum_{i \neq j}^{\ast} \mathrm{tr}(\mathbf X_{i}\mathbf X^T_{i} \circ \mathbf X_{j}\mathbf X^T_{j})-2\frac{1}{P^N_3}\sum_{i \neq j\neq k}^{\ast} \mathrm{tr}(\mathbf X_{i}\mathbf X^T_{i} \circ \mathbf X_{j}\mathbf X^T_{k})+\frac{1}{P^N_4} \sum_{i\neq j \neq k \neq l}^{\ast} \mathrm{tr}(\mathbf X_{i}\mathbf X^{T}_{j} \circ \mathbf X_{k}\mathbf X^{T}_{l})
\end{align*}
is an unbiased and ratio-consistent estimator to $\mathrm{tr}(\mathbf D^2_{\boldsymbol \Sigma})$. The construction of $Y_{3N}$ is closely related to that of $Y_{1N}$ and $Y_{2N}$. To see this, note that $Y_{3N}$ is a linear combination of three $U$-statistics, $U_{3N}$, the unbiased estimator of $\mathrm{tr}(\mathbf D^2_{\boldsymbol \Sigma})$ when $\boldsymbol \mu=\mathbf 0$, and $U_{7N}$ and $U_{8N}$, which make the bias of $Y_{3N}$ zero when $\boldsymbol \mu \neq \mathbf 0$. Hence, a consistent estimator of $\lambda_{D}$ is
\begin{equation*}
\hat{\lambda}_{D}=\frac{Y_{2N}+Y^2_{1N}-2Y_{3N}}{NY_{2N}+Y^2_{1N}-(N+1) Y_{3N}}.
\end{equation*}

\subsection{Remarks}
There is no need to account for the mean vector when the random vectors are centered. In this case, the proposed shrinkage covariance matrix estimators can be obtained by replacing $N-1$ with $N$ in the formula for $\hat{\lambda}$, $\hat{\lambda}_I$ or $\hat{\lambda}_D$, the sample covariance matrix $\mathbf S$ with $\sum_{i=1}^p \mathbf X_i \mathbf X_i^{T}/N$ and the statistics $Y_{1N}$, $Y_{2N}$ and $Y_{3N}$ with $U_{1N}$, $U_{2N}$ and $U_{3N}$ respectively. The last modification is due to the fact that $U_{1N}$, $U_{2N}$ and $U_{3N}$ are unbiased and ratio-consistent estimators to the targeted parameters when $\boldsymbol \mu =\mathbf 0$. 

\indent \cite{Fisher2011} derived Stein-type shrinkage covariance matrix estimators for the three target matrices considered herein under a multivariate normal model. We emphasize that our estimators differ in three important aspects. First, the consistency of the proposed estimators for the optimal shrinkage intensities $\lambda$, $\lambda_I$ or $\lambda_D$ is not sensitive to departures of the normality assumption. Second, $\mathrm{tr}(\boldsymbol \Sigma^2)$ and $\mathrm{tr}(\mathbf D_{\boldsymbol \Sigma})$ are estimated using $U$-statistics and are not based on the sample covariance matrix $\mathbf S$ as in \cite{Fisher2011}. Consequently, we avoid terms such as $(\mathbf X^{T}_{i}\mathbf X_{i})^2$ or $\mathbf X_{ia}^4$, which allows us to control their asymptotic variance. Third, the class of covariance matrices under consideration in \cite{Fisher2011} is different. They require the first four arithmetic means of the eigenvalues of $\boldsymbol \Sigma$ to converge while we place trace ratios restrictions on $\boldsymbol \Sigma$.

\subsection{Software implementation}
The R \citep{RCoreTeam2013} language package \textit{Shrinkcovmat} implements the proposed Stein-type shrinkage covariance estimators and it is available at \href{http://cran.r-project.org/web/packages/ShrinkCovMat}{http://cran.r-project.org/web/packages/ShrinkCovMat}. The core functions \textit{shinkcovmat.equal}, \textit{shinkcovmat.identity} and \textit{shinkcovmat.unequal} provide the proposed shrinkage covariance matrix estimators when $\mathbf T=\nu \mathbf I_p$, $\mathbf T=\mathbf I_p$ and $\mathbf T= \mathbf D_{\mathbf S}$ respectively. The statistics $Y_{1N}$, $Y_{2N}$ and $Y_{3N}$ are calculated using the computationally efficient formulas given in \ref{Appendix3}. To modify the shrinkage estimators when $\boldsymbol \mu=\mathbf 0$, one should set the argument \textit{centered}=TRUE in the core functions. 

\section{Simulations}\label{Sec03}
We carried out a simulation study to investigate the performance of the proposed Stein-type covariance matrix estimators. The $p$-variate random vectors $\mathbf X_1$,$
\ldots$,$\mathbf X_N$ were generated according to model~(\ref{Nonparametricmodel}), where we employed the following three distributional scenarios regarding $\mathbf Z_1,\ldots,\mathbf Z_N$: 
\begin{enumerate}
\item A normality scenario, in which $Z_{ia} \stackrel{i.i.d}{\sim} \mathrm{N}(0,1)$.
\item A gamma scenario, in which $Z_{ia}=(Z^{\ast}_{ia}-8)/4$, $Z^{\ast}_{ia} \stackrel{i.i.d}{\sim} \mathrm{Gamma}(4,0.5)$ and thus $B=12$.  
\item A mixture of Scenarios 1 and 2, in which the first $p/2$ elements of $\mathbf Z_i$ are distributed according to Scenario 1 and the remaining elements according to Scenario 2.
\end{enumerate}
Scenarios 2 and 3 were used to empirically verify the nonparametric nature of the proposed methodology. To mimic high-dimensional settings, we let $N$ range from $10$ to $100$ with increments of $10$ and we let $p=100,500,1000,1500$ and $2500$. The proposed family of covariance estimators was evaluated using $\hat{\mathbf S}^{\star}$, the proposed shrinkage covariance matrix estimator when $\mathbf T=\nu \mathbf I_p$, which was then compared to $\hat{\mathbf S}^{\star}_{LW}$ and $\hat{\mathbf S}^{\star}_{FS}$, the corresponding shrinkage covariance matrix estimator proposed by \cite{Ledoit2004} and \cite{Fisher2011} respectively. As in \cite{Ledoit2004}, we assumed that $\boldsymbol \mu=\mathbf 0$ and we made the necessary adjustments to $\hat{\mathbf S}^{\star}$ and $\hat{\mathbf S}^{\star}_{FS}$. Since $\hat{\mathbf S}^{\star}_{FS}$ was constructed under a multivariate normal model assumption, its performance was evaluated only in sampling schemes that involved Scenario 1. We excluded the shrinkage estimator of \cite{Schafer2005} in our simulation studies because they do not guarantee consistent estimation of $\lambda$ in high-dimensional settings. Given $\boldsymbol \Sigma$, $N$, $p$ and the distributional scenario, we draw $1000$ replicates based on which we calculated the simulated percentage relative improvement in average loss (SPRIAL) of $\hat{\mathbf S}^{\star}$ and of $\hat{\mathbf S}^{\star}_{FS}$ for estimating $\boldsymbol \Sigma$. Formally, the SPRIAL criterion of $\hat{\boldsymbol \Sigma}$ was defined as
$$\mathrm{SPRIAL}(\hat{\boldsymbol \Sigma})=\frac{\sum_{b=1}^{1000}||\hat{\mathbf S}^{\star}_{LW,b}-\boldsymbol \Sigma||^2_F-\sum_{b=1}^{1000}||\hat{\boldsymbol \Sigma}_{b}-\boldsymbol \Sigma||^2_F}{\sum_{b=1}^{1000}||\hat{\mathbf S}^{\star}_{LW,b}-\boldsymbol \Sigma||^2_F}\times 100\%,$$
where $\hat{\mathbf S}^{\star}_{LW,b}$ denotes the estimator of \cite{Ledoit2004} at replicate $b$ ($b=1,\ldots,1000$) and $\hat{\boldsymbol \Sigma}_{b}$ denotes the corresponding estimator of the competing estimation process that generates the covariance matrix estimator $\hat{\boldsymbol \Sigma}$. By definition, $\mathrm{SPRIAL}(\boldsymbol \Sigma)=100\%$ and $\mathrm{SPRIAL}(\hat{\mathbf S}^{\star}_{LW})=0\%$. Therefore, positive (negative) values of $\mathrm{SPRIAL}(\hat{\boldsymbol \Sigma})$ imply that $\hat{\boldsymbol \Sigma}$ is a more (less) efficient covariance matrix estimator than $\hat{\mathbf S}^{\star}_{LW}$ while values around zero imply that the two estimators were equally efficient. Note that we treated $\hat{\mathbf S}^{\star}_{LW}$ as the baseline estimator because \cite{Ledoit2004} have already established its efficiency over the sample covariance matrix $\mathbf S$. 

\begin{figure}[!bt] 
\centering{
\includegraphics[scale=0.4]{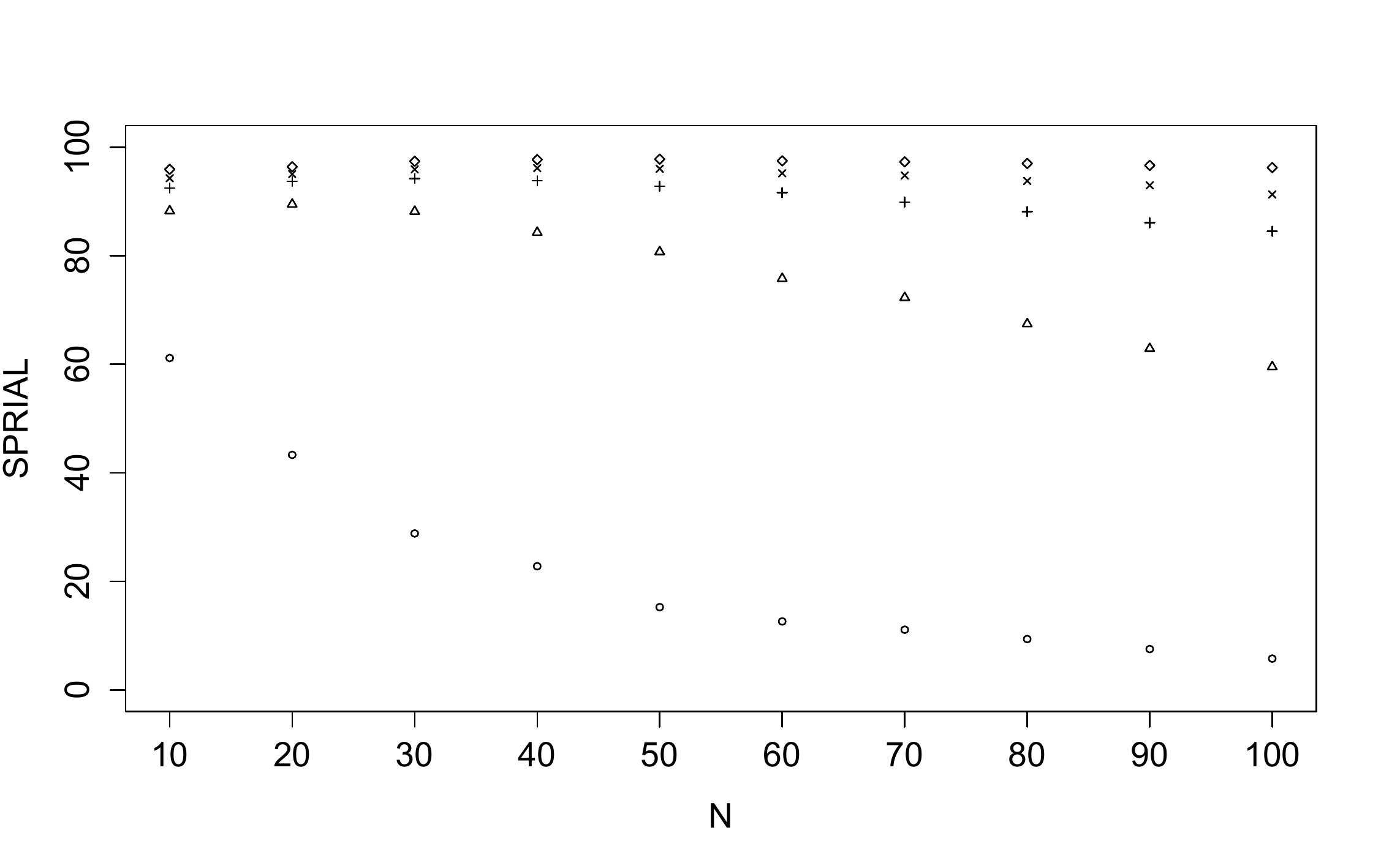}}
\caption{$\mathrm{SPRIAL}(\hat{\mathbf S}^{\star})$ under Scenario 2 when $\boldsymbol \Sigma=\mathbf I_p$ and $p=100$ ($\circ$ symbols), $500$ ($+$ symbols), $1000$ ($\triangle$ symbols), $1500$ (x symbols) or $2500$ ($\diamond$ symbols).}
\label{Fig01}
\end{figure}

\indent To explore the situation in which the target matrix equals the true covariance matrix, we set $\boldsymbol \Sigma=\mathbf I_p$. In this case, accurate estimation of $\lambda$ is crucial due to the fact that $\lambda=1$ regardless of $N$ and $p$. Table~\ref{sim01} contains the simulation results under Scenario 1 for the mean of the estimated optimal shrinkage intensities based on the proposed method ($\hat{\lambda}$), the approach of \cite{Ledoit2004} ($\hat{\lambda}_{LW}$) and the approach of \cite{Fisher2011} ($\hat{\lambda}_{FS}$) when $N=10,50,100$ and $p=100,1000,2500$. As required, $\hat{\lambda}$, $\hat{\lambda}_{LW}$ and $\hat{\lambda}_{FS}$ were all restricted to lie on the unit interval. Compared to $\hat{\lambda}_{LW}$, $\hat{\lambda}$ appeared to be more accurate in estimating $\lambda$ as it was substantially less biased and with slightly smaller standard error for all $N$ and $p$. Although the bias of $\hat{\lambda}_{LW}$ decreased as $N$ increased, $\hat{\lambda}_{LW}$ seemed to be biased downwards even when $N=100$. These trends also occurred for Scenarios 2 and 3, and for this reason the results are omitted. As expected, no significant difference was noticed between $\hat{\lambda}$ and $\hat{\lambda}_{FS}$ under Scenario 1. Figure~\ref{Fig01} displays $\mathrm{SPRIAL}(\hat{\mathbf S}^{\star})$ under Scenario 2 - similar patterns were observed for the other two distributional scenarios. Clearly, $\hat{\mathbf S}^{\star}$ was more effective than $\hat{\mathbf S}^{\star}_{LW}$ for $p \leq 500$ ($59.54\%-97.81\%$) and for $p=100$ and $N \leq 50$ ($15.24\%-61.16\%$). We should mention that $\hat{\mathbf S}^{\star}$ and $\hat{\mathbf S}^{\star}_{FS}$ were equally efficient, meaning that $\mathrm{SPRIAL}(\hat{\mathbf S}^{\star})$ and $\mathrm{SPRIAL}(\hat{\mathbf S}^{\star}_{FS})$ took similar values.

\begin{table}[!tb]
\centering
\caption{Estimation of the optimal shrinkage intensity $\lambda=1$ under Scenario 1 when $\boldsymbol \Sigma=\mathbf I_p$.}
\begin{tabular*}{\textwidth}{c @{\extracolsep{\fill}}ccccccccc}
\toprule
  &     & \multicolumn{2}{c}{\small{$\hat{\lambda}$}} & \multicolumn{2}{c}{$\hat{\lambda}_{LW}$} & \multicolumn{2}{c}{$\hat{\lambda}_{FS}$}\\
		\cline{3-4} \cline{5-6} \cline{7-8} 
  $N$ & $p$ & Mean & S.E. & Mean & S.E. & Mean & S.E.\\ 
  \midrule
  10 & 100   & 0.9914 & 0.0133 & 0.8997 & 0.0196 & 0.9925 & 0.0119 \\ 
     & 1000  & 0.9992 & 0.0013 & 0.9000 & 0.0019 & 0.9992 & 0.0012 \\ 
     & 2500  & 0.9997 & 0.0005 & 0.9000 & 0.0008 & 0.9997 & 0.0005 \\ 
  50 & 100   & 0.9924 & 0.0113 & 0.9789 & 0.0167 & 0.9925 & 0.0112 \\ 
     & 1000  & 0.9992 & 0.0012 & 0.9800 & 0.0020 & 0.9992 & 0.0012 \\ 
     & 2500  & 0.9997 & 0.0005 & 0.9800 & 0.0008 & 0.9997 & 0.0005 \\ 
 100 & 100   & 0.9923 & 0.0114 & 0.9864 & 0.0145 & 0.9924 & 0.0113 \\ 
     & 1000  & 0.9992 & 0.0012 & 0.9900 & 0.0020 & 0.9992 & 0.0011 \\ 
     & 2500  & 0.9997 & 0.0005 & 0.9900 & 0.0008 & 0.9997 & 0.0004 \\ 
\bottomrule
\end{tabular*}
\label{sim01}
\end{table}

\begin{figure}[!tb] \centering{
\includegraphics[scale=0.4]{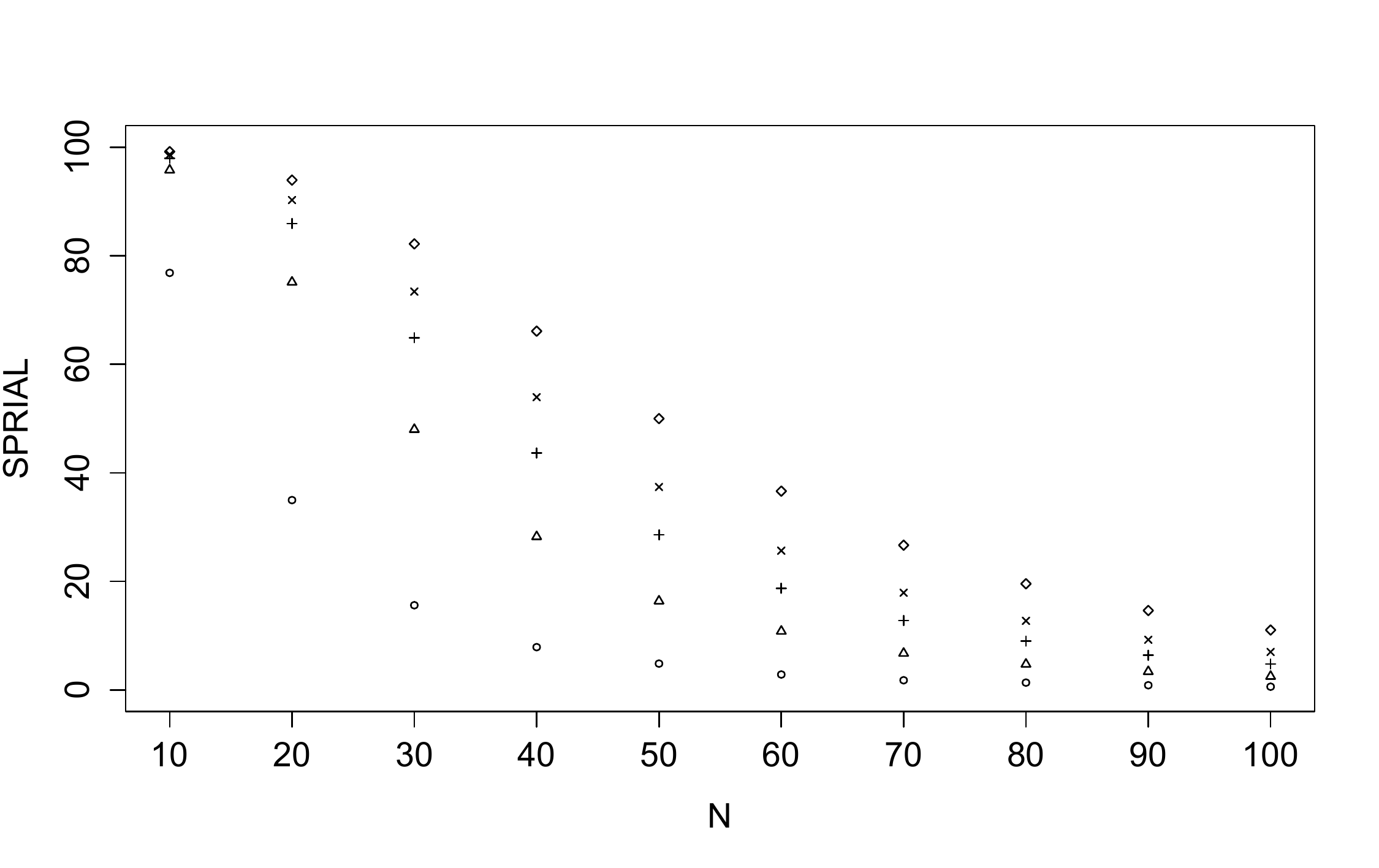}}
\caption{$\mathrm{SPRIAL}(\hat{\mathbf S}^{\star})$ under Scenario 3 when $\boldsymbol \Sigma$ is a tridiagonal correlation matrix with $(a,b)$-th element $\Sigma_{ab}=0.1 I(|a-b|)$ and $p=100$ ($\circ$ symbols), $500$ ($+$ symbols), $1000$ ($\triangle$ symbols), $1500$ (x symbols) or $2500$ ($\diamond$ symbols).}
\label{Fig02}
\end{figure}

\indent To investigate the performance of $\hat{\mathbf S}^{\star}$ when $\boldsymbol \Sigma$ deviates slightly from the target matrix, we employed a tridiagonal correlation matrix for $\boldsymbol \Sigma$ in which the non-zero off-diagonal elements were all equal to $0.1$. Figure~\ref{Fig02} suggests that $\hat{\mathbf S}^{\star}$ outperformed $\hat{\mathbf S}^{\star}_{LW}$ in extreme high-dimensional settings ($N\leq 50$) under Scenario 3. The efficiency gains in using $\hat{\mathbf S}^{\star}$ instead of $\hat{\mathbf S}^{\star}_{LW}$ decreased at a much faster rate than before as $N$ increased. Note that similar trends were observed under Scenarios 1 and 2 (results not shown). 
 
\begin{figure}[!tb] 
\centering{
\includegraphics[scale=0.4]{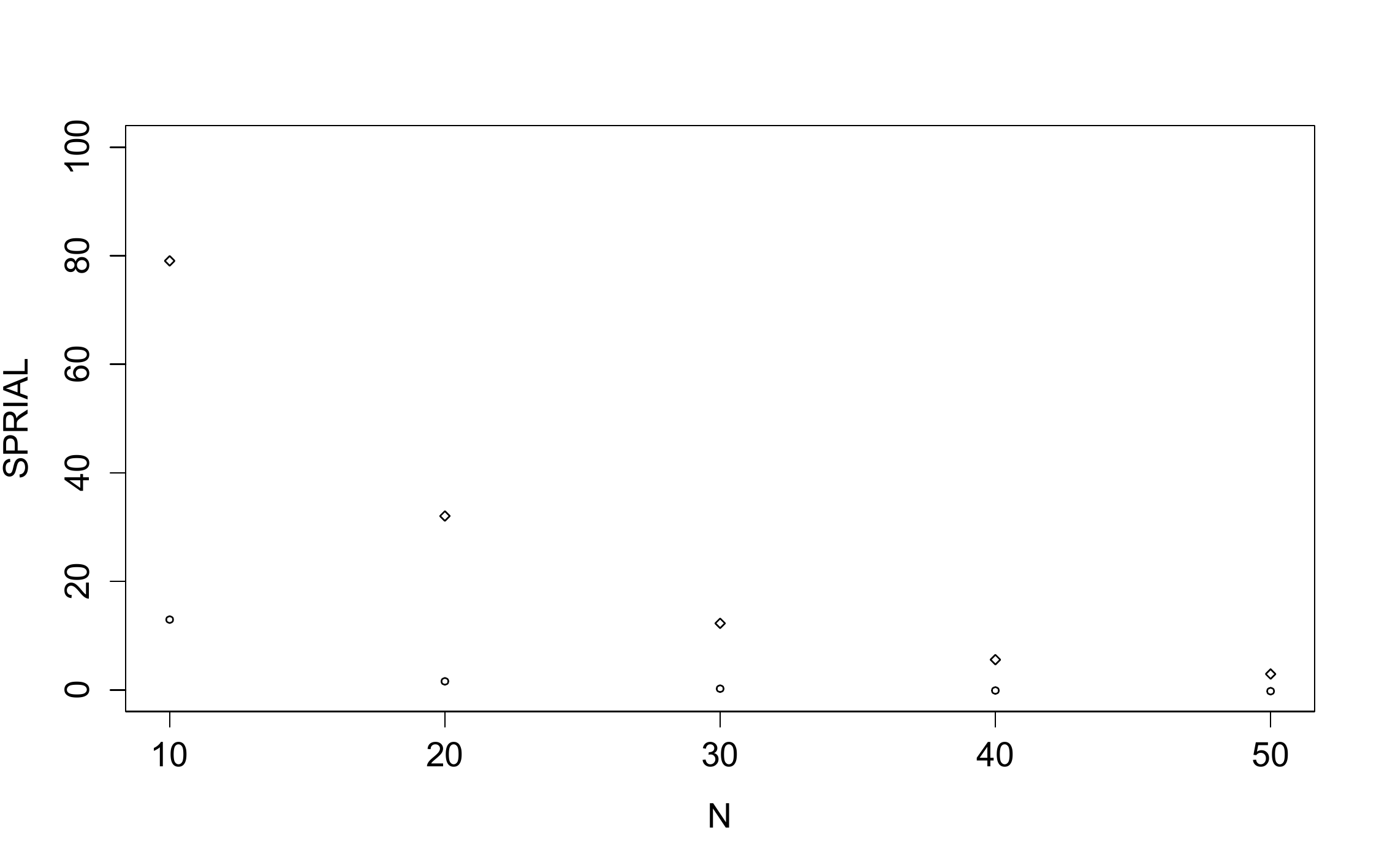}}
\caption{$\mathrm{SPRIAL}(\hat{\mathbf S}^{\star})$ under Scenario 2 when $\boldsymbol \Sigma$ satisfies a first-order autoregressive form with correlation parameter $\rho=0.5$ and $p=100$ ($\circ$ symbols) or $2500$ ($\diamond$ symbols).}
\label{Fig03}
\end{figure}

\indent Next, we considered dependence structures such that $\nu \mathbf I_p$ is a rather poor approximation of $\boldsymbol \Sigma$. First, we defined $\boldsymbol \Sigma$ as a first order autoregressive correlation matrix with $(a,b)$-th element $\Sigma_{ab}=0.5^{|a-b|}$. Although the results in Table~\ref{sim02} imply that $\hat{\lambda}$ is a more accurate estimator of $\lambda$ than $\hat{\lambda}_{LW}$, Figure~\ref{Fig03} suggests that $\hat{\mathbf S}^{\star}$ and $\hat{\mathbf S}^{\star}_{LW}$ were almost equally efficient as soon as $N=30$ and regardless of the dimensionality. We reached the same conclusions when we let $\boldsymbol \Sigma$ to be either a positive definite matrix whose $p$ eigenvalues were drawn from the uniform distribution $\mathrm{U}(0.5,10)$ or a block diagonal covariance matrix such that the dimension of each block matrix is $p/4 \times p/4$ and the eigenvalues of these 4 block matrices were drawn from $\mathrm{U}(0.5,5)$, $\mathrm{U}(5,10)$, $\mathrm{U}(10,20)$ and $\mathrm{U}(0.5,100)$ respectively. In particular, $\hat{\lambda}$ seemed to outperformed $\hat{\lambda}_{LW}$ for all configurations of $(N,p,\boldsymbol \Sigma)$ while $\mathrm{SPRIAL}(\hat{\mathbf S}^{\star})$ was close to zero unless $N \leq 30$. Hence, if the target matrix $\nu \mathbf I_p$ fails to describe adequately the dependence structure, we expect $\hat{\mathbf S}^{\star}$ to be more efficient than $\hat{\mathbf S}^{\star}_{LW}$ only for small sample sizes.

\indent Further, we adopted a compound symmetry correlation form for $\boldsymbol \Sigma$ with correlation parameter $\rho=0.5$. This is the only configuration of $\boldsymbol \Sigma$ where $t_1 \neq 0 $ and $t_2 \neq 0$ in assumption~(\ref{CovMatAss}). The estimator $\hat{\lambda}_{LW}$ appeared to be less biased but with larger standard error than $\hat{\lambda}$, while the $\mathrm{SPRIAL}(\hat{\mathbf S}^{\star})$ was always close to zero. The performance of $\hat{\mathbf S}^{\star}$ and $\hat{\mathbf S}^{\star}_{LW}$ was comparable across the related sampling schemes.

\begin{table}[!tb]
\centering
\caption{Estimation of $\lambda$ under Scenario 1 when $\boldsymbol \Sigma$ satisfies a first-order auto-regressive correlation form with correlation parameter $\rho=0.5$.}
\begin{tabular*}{\textwidth}{c @{\extracolsep{\fill}}ccccccccc}
  \toprule
	   &     &          & \multicolumn{2}{c}{\small{$\hat{\lambda}$}} & \multicolumn{2}{c}{$\hat{\lambda}_{LW}$} & \multicolumn{2}{c}{$\hat{\lambda}_{FS}$}\\
\cline{4-5} \cline{6-7} \cline{8-9}
 $N$ & $p$ &$\lambda$ & Mean & S.E. & Mean & S.E. & Mean & S.E.\\ 
  \midrule
10 & 100 & 0.9392 & 0.9418 & 0.0300 & 0.8474 & 0.0277 & 0.9422 & 0.0277 \\ 
   & 1000 & 0.9934 & 0.9933 & 0.0035 & 0.8940 & 0.0032 & 0.9933 & 0.0031 \\ 
   & 2500 & 0.9973 & 0.9973 & 0.0014 & 0.8976 & 0.0013 & 0.9973 & 0.0013 \\ 
  50 & 100 & 0.7556 & 0.7571 & 0.0240 & 0.7418 & 0.0237 & 0.7571 & 0.0237 \\ 
   & 1000 & 0.9678 & 0.9676 & 0.0033 & 0.9482 & 0.0032 & 0.9675 & 0.0032 \\ 
   & 2500 & 0.9869 & 0.9868 & 0.0014 & 0.9671 & 0.0013 & 0.9868 & 0.0013 \\ 
  100 & 100 & 0.6071 & 0.6080 & 0.0175 & 0.6018 & 0.0175 & 0.6080 & 0.0174 \\ 
   & 1000 & 0.9377 & 0.9375 & 0.0032 & 0.9281 & 0.0032 & 0.9375 & 0.0032 \\ 
   & 2500 & 0.9741 & 0.9741 & 0.0013 & 0.9643 & 0.0013 & 0.9741 & 0.0013 \\ 
   \bottomrule
\end{tabular*}
\label{sim02}
\end{table}

\begin{figure}[!tb] 
\centering{
\includegraphics[scale=0.4]{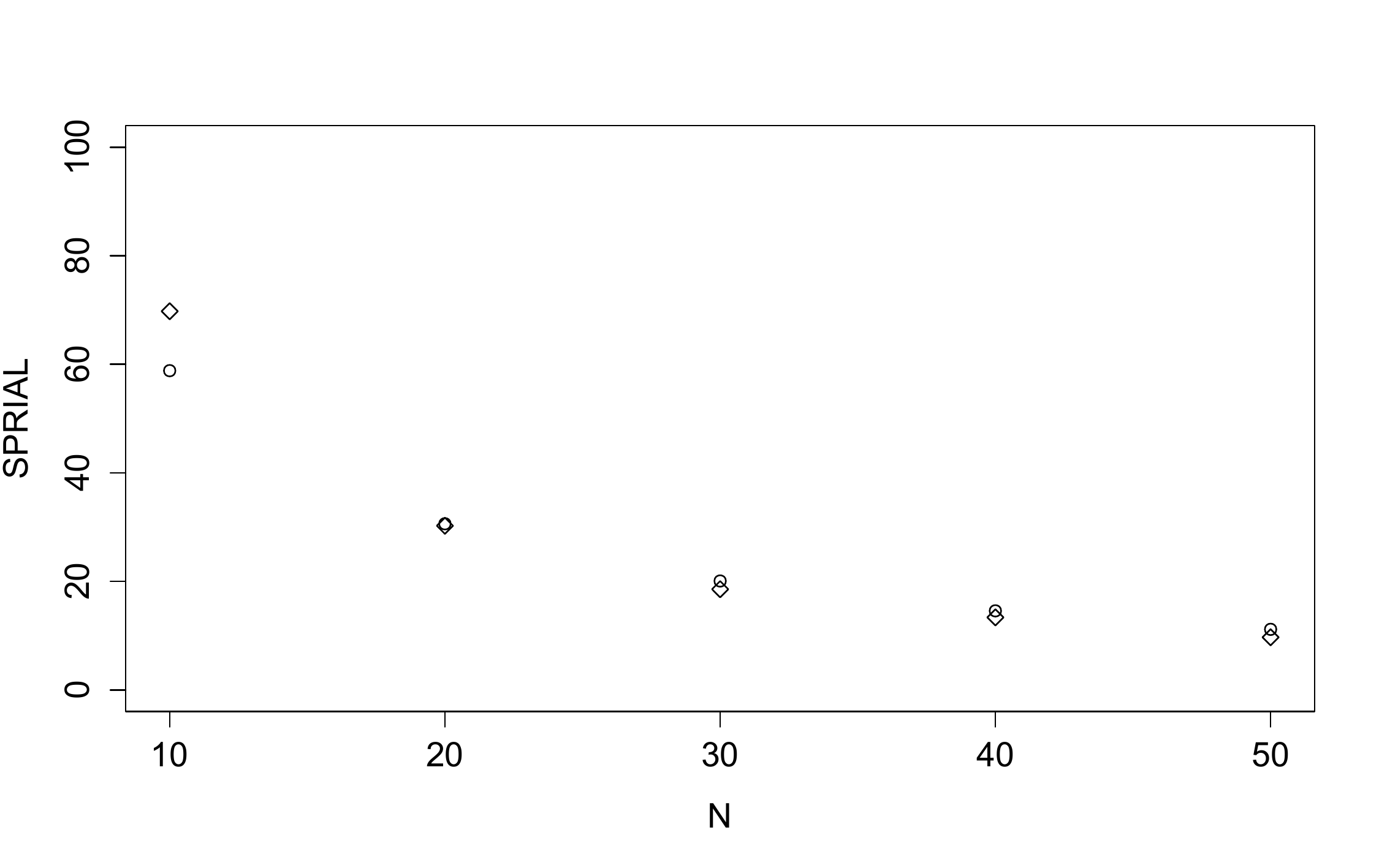}}
\caption{SPRIAL criterion for the inverse of $\hat{\mathbf S}^{\star}$ under Scenario 3 when $\boldsymbol \Sigma$ satisfies a compound symmetry correlation matrix and $p=100$ ($\circ$ symbols) or $2500$ ($\diamond$ symbols).}
\label{Fig04}
\end{figure}

\indent We also evaluated the performance of the inverse of the competing Stein-type shrinkage covariance matrix estimators using a modified version of the $\mathrm{SPRIAL}$ criterion, that is we replaced $\mathbf \Sigma$, $\widehat{\mathbf \Sigma}$ and $\hat{\mathbf S}^{\star}_{LW}$ with their corresponding inverses in $\mathrm{SPRIAL}(\hat{\boldsymbol \Sigma})$. The inverse of $\hat{\mathbf S}^{\star}$ was more efficient that of $\hat{\mathbf S}^{\star}_{LW}$ when $\boldsymbol \Sigma$ was equal to the identity matrix or to the tridiagonal correlation matrix, and quite surprisingly when $\boldsymbol \Sigma$ satisfied the compound symmetry correlation pattern and $N \leq 50$, as shown in Figure~\ref{Fig04}. In all other sampling schemes, the inverses of $\hat{\mathbf S}^{\star}$ and $\hat{\mathbf S}^{\star}_{LW}$ were comparable. 

\indent Based on our simulations, $\mathrm{SPRIAL}(\hat{\mathbf S}^{\star})$ was an increasing function of $p$ and a decreasing function of $N$ while keeping the remaining parameters fixed. This indicates that, compared to $\hat{\mathbf S}^{\star}_{LW}$, $\hat{\mathbf S}^{\star}$ is an improved estimator of $\boldsymbol \Sigma$ in extreme high-dimensional settings. The nonparametric nature of $\hat{\mathbf S}^{\star}$ was empirically verified because both the SPRIAL criterion and the bias of $\hat{\lambda}$ seemed to remain constant across the three distributional scenarios. In addition, we find reassuring that the simulation results for $\hat{\lambda}$ and $\hat{\lambda}_{FS}$ and for $\hat{\mathbf S}^{\star}$ and $\hat{\mathbf S}^{\star}_{FS}$ were almost identical under a multivariate normal distribution. To this end, we also believe that the above trends together with the fact that we have not encountered a situation in which $\hat{\mathbf S}^{\star}$ was performing significantly worse than $\hat{\mathbf S}^{\star}_{LW}$ are favoring the consistency of the proposed covariance matrix estimators.

\section{Empirical Study: Colon Cancer Study}\label{Sec04}
\begin{figure}[!tb]
\centering
\includegraphics[scale=0.5]{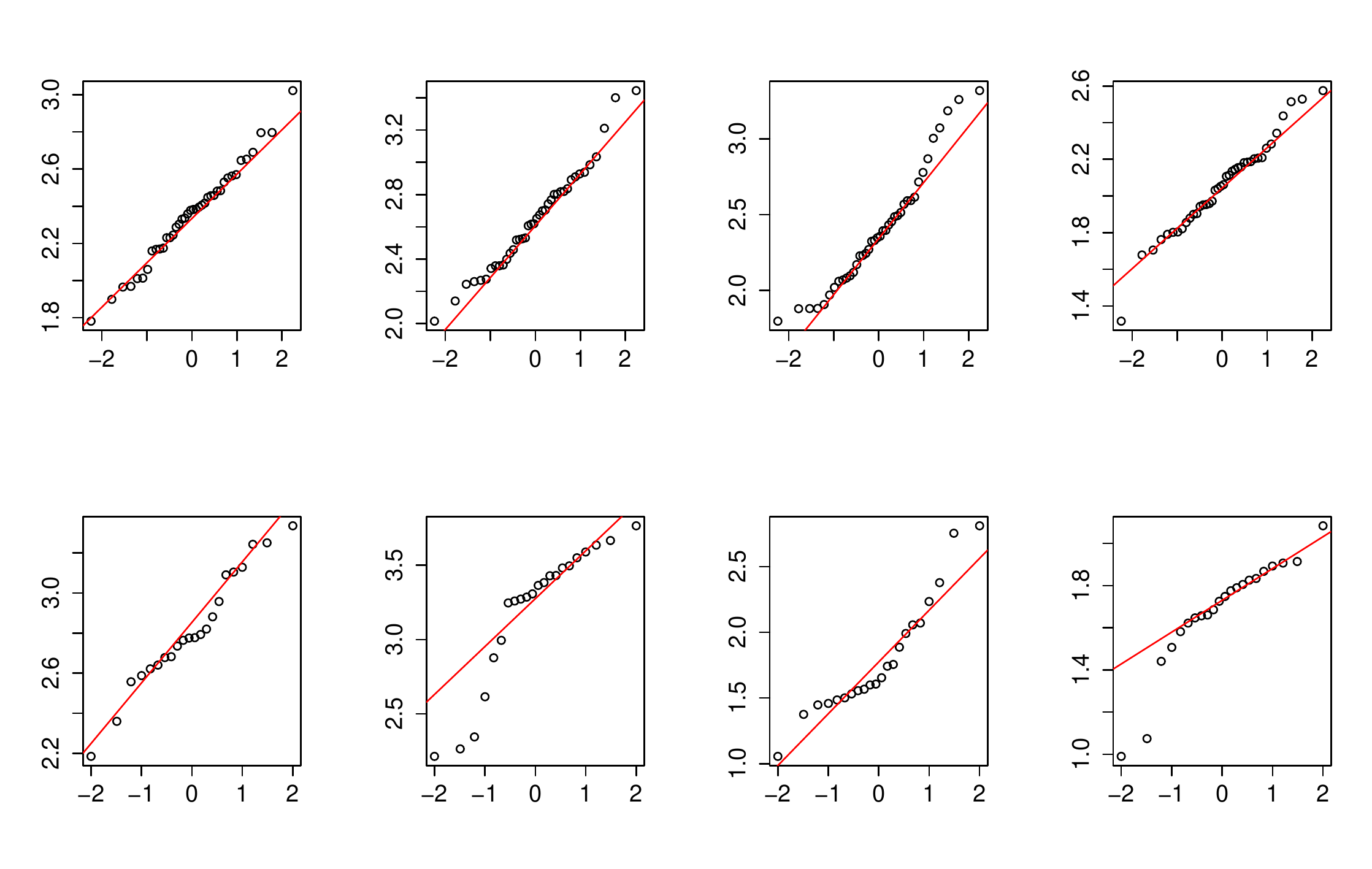}
\caption{Quantile-Quantile plots for the expression levels of the 4 top genes according to the BW selection criterion. The top panel corresponds to the normal group and the bottom panel to the colon cancer group.}
\label{Fig1}
\end{figure}

\cite{Alon1999} described a colon cancer study where expression levels for $2000$ genes were measured on $40$ normal and on $22$ colon tumor tissues. The dataset is available at \href{http://genomics-pubs.princeton.edu/oncology/affydata}{http://genomics-pubs.princeton.edu/oncology/affydata}. As in \cite{Fisher2011}, we apply a logarithmic (base 10) transformation to the expression levels and sort the genes based on the between-group to within-group sum of squares (BW) selection criterion \citep{Dudoit2002}. In the literature, it has been suggested to estimate the covariance matrix of the genes using a subset of the $p$ top genes \citep[see, e.g.,][]{Fisher2011}. With this in mind, we plan to estimate the covariance matrix of the normal and of the colon cancer group for subsets of the top $p$ genes, where $p=250$, $500$, $750$, $1000$, $1250$, $1500$, $1750$ and $2000$. The Quantile-Quantile plots in Figure~\ref{Fig1} raise concerns regarding the assumption that the 4 top genes are marginally normally distributed. Since similar patterns occur for the remaining genes, we conclude that a multivariate normal model is not likely to hold and nonparametric covariance matrix estimation is required. For this purpose, we calculate $\widehat{\mathbf S}^{\star}$, $\widehat{\mathbf S}^{\star}_I$ and $\widehat{\mathbf S}^{\star}_D$ for both groups and for all values of $p$.

\begin{table}
\centering
\caption{The estimates of $\lambda$, $\lambda_I$, $\lambda_D$, $\nu$ and the range of the $p$ sample variances ($r$) in the colon cancer dataset.}
\label{tab1}
\begin{tabular*}{\textwidth}{c @{\extracolsep{\fill}} cccccccccc}
  \toprule
  &   &  \multicolumn{8}{c}{$p$}\\
	\cline{3-10}
Group & Statistic & 250 & 500 & 750 & 1000 & 1250 & 1500 & 1750 & 2000 \\ 
\midrule
Normal & $\hat{\lambda}$       & 0.1407 & 0.1467 & 0.1465 & 0.1454 & 0.1435 & 0.1423 & 0.1414 & 0.1401 \\ 
       & $\hat{\nu}$           & 0.0999 & 0.0963 & 0.0938 & 0.0916 & 0.0902 & 0.0894 & 0.0889 & 0.0882 \\ 
       & $\hat{\lambda}_D$     & 0.1402 & 0.1464 & 0.1463 & 0.1452 & 0.1434 & 0.1422 & 0.1413 & 0.1400 \\ 
       & $r$                   & 0.4604 & 0.4638 & 0.4700 & 0.4714 & 0.4714 & 0.4714 & 0.4714 & 0.4714 \\ 
       & $\hat{\lambda}_I$     & 0.0564 & 0.0791 & 0.0913 & 0.0987 & 0.1036 & 0.1075 & 0.1105 & 0.1125 \\ 
 \midrule
Colon  & $\hat{\lambda}$       & 0.2035 & 0.2048 & 0.1970 & 0.1959 & 0.1952 & 0.1967 & 0.1969 & 0.1956 \\ 
  & $\hat{\nu}$                & 0.1113 & 0.1060 & 0.1033 & 0.0996 & 0.0984 & 0.0975 & 0.0965 & 0.0958 \\ 
  & $\hat{\lambda}_D$          & 0.2027 & 0.2044 & 0.1967 & 0.1957 & 0.1950 & 0.1966 & 0.1968 & 0.1955 \\ 
  & $r$                        & 0.4107 & 0.4107 & 0.4201 & 0.4201 & 0.4226 & 0.4226 & 0.4226 & 0.4226 \\ 
  & $\hat{\lambda}_I$          & 0.1081 & 0.1367 & 0.1476 & 0.1542 & 0.1599 & 0.1654 & 0.1688 & 0.1705 \\ 
   \bottomrule
\end{tabular*}
\end{table}

\indent The optimal shrinkage intensity $\lambda_T$ in (\ref{bfmatrix}) is informative for the suitability of the selected target matrix $\mathbf T$ because it reveals its contribution to $\mathbf S_T^{\star}$. If $\hat{\lambda}$, $\hat{\lambda}_I$ and $\hat{\lambda}_D$ differ significantly, then it is meaningful to choose the target matrix with the largest estimated optimal shrinkage intensity. Otherwise, the selection of the target matrix can be based on $\hat{\nu}$ and $r$, the range of the $p$ sample variances. In particular, we suggest to employ $\mathbf I_p$ when $\hat{\nu}$ is close to $1.00$, $\mathbf D_{\mathbf S}$ when $r$ is large (say more than one unit so as to account for the sampling variability), and $\nu \mathbf I_p$ when neither of these seems plausible. Table~\ref{tab1} displays the estimated optimal shrinkage intensities for the proposed family of shrinkage covariance estimators, $\hat{\nu}$ and $r$ in both groups. The estimates of $\lambda$ and $\lambda_D$ are very similar in both groups for all subsets of genes and larger than those of $\lambda_I$. This suggests that $\mathbf D_{\mathbf S}$ and $\nu\mathbf I_p$ are better options than $\mathbf I_p$ as a target matrix. Since $r$ appears to be relatively small and constant across the top $p$ genes in both groups, it seems sensible to set $\mathbf T=\nu\mathbf I_p$ for all $p$. Therefore, we recommend using $\widehat{\mathbf S}^{\star}$ to estimate the covariance matrix of the genes in the normal and in the colon cancer group and regardless of $p$.

\section{Discussion}\label{Sec05}
We proposed a new family of nonparametric Stein-type shrinkage estimators for a high-dimensional covariance matrix. This family is based on improving the nonparametric estimation of the optimal shrinkage intensity for three commonly used target matrices at the expense of imposing mild restrictions for the covariance matrix $\boldsymbol \Sigma$. In our simulations, the proposed shrinkage covariance matrix estimator $\hat{\mathbf S}^{\star}$ was more precise than that of \cite{Ledoit2004} for estimating $\mathbf \Sigma$, especially when the number of variables $p$ is extremely large compared to the sample size $N$ and/or $\nu\mathbf I_p$ is a good approximation of $\mathbf \Sigma$. Unsurprisingly, the behavior of our estimators and that of \cite{Fisher2011} were similar as long as an underlying multivariate normal model holds. However, we emphasize that our estimators are more flexible since they are robust to departures from normality. In addition, we recommended a simple data-driven strategy for selecting the target matrix in~(\ref{bfmatrix}). The main idea is to compare $\hat{\lambda},\hat{\lambda}_I$ and $\hat{\lambda}_D$ and choose the target matrix with the largest estimated optimal shrinkage intensity. If these estimates are similar and $\hat{\nu}$ is close to one, then it is sensible to use $\mathbf I_p$ as target matrix. Otherwise, $\nu\mathbf I_p$ should be selected when the $p$ sample variances are very close and $\mathbf D_\mathbf{S}$ when these vary. The R package \textit{ShrinkCovMat} implements the proposed estimators. 

\indent Since $\widehat{\mathbf S}^{\star}$, $\widehat{\mathbf S}^{\star}_I$ and $\widehat{\mathbf S}^{\star}_D$ are all biased estimators of $\boldsymbol \Sigma$, their risk functions might be unbounded especially when none of the corresponding target matrices describes adequately the underlying dependence structure. In these situations, a more suitable target matrix $T$ in~(\ref{bfmatrix}) must be considered. The corresponding $\lambda_{T}$ should be estimated by following our guidelines in Section~\ref{AltMat}. If such a target matrix cannot be identified, one should choose among the alternative covariance matrix estimation methods mentioned in the Introduction. 

\indent In future research, we aim to investigate formal procedures for selecting the target matrix in~(\ref{bfmatrix}) and to extend the proposed Steinian shrinkage approach for estimating correlation matrices.

\appendix \label{Appendix}
\section{Useful Results}\label{Appendix1}
We list six results with respect to model~(\ref{Nonparametricmodel}) that allows us to derive the formulas for the $\lambda$, $\lambda_I$ and $\lambda_D$:
\begin{enumerate}
\item $\mathrm{E}[\mathrm{tr}(\mathbf S)]=\mathrm{tr}(\boldsymbol \Sigma)$.
\item $\mathrm{E}[\mathrm{tr}(\mathbf S^2)]=\frac{N}{N-1}\mathrm{tr}(\boldsymbol \Sigma^2)+\frac{1}{N-1}\mathrm{tr}^2(\boldsymbol \Sigma)+\frac{B}{N-1} \mathrm{tr}(\mathbf D^{2}_{\boldsymbol \Sigma})$.
\item $\mathrm{E}[\mathrm{tr}^2(\mathbf S)]=\mathrm{tr}^2(\boldsymbol \Sigma)+\frac{2}{N-1}\mathrm{tr}(\boldsymbol \Sigma^2)+\frac{B}{N-1} \mathrm{tr}(\mathbf D^{2}_{\boldsymbol \Sigma})$.
\item $\mathrm{E}[\mathrm{tr}(\mathbf D_{\mathbf S})]=\mathrm{tr}(\mathbf D_{\boldsymbol \Sigma})$.
\item $\mathrm{E}[\mathrm{tr}(\mathbf D^{2}_{\mathbf S})]=\mathrm{E}[\mathrm{tr}(\mathbf S\mathbf D_{\mathbf S})]=\frac{N+1}{N-1}\mathrm{tr}(\mathbf D^2_{\boldsymbol \Sigma})+\frac{B}{N-1}\sum_{a=1}^{p}\sum_{b=1}^{p} \left(\Sigma^{1/2}_{ab}\right)^{4}$, where $\Sigma^{1/2}_{ab}$ denotes the $(a,b)$-th element of $\boldsymbol \Sigma^{1/2}$.
\item $\mathrm{E}[\mathrm{tr}(\mathbf \Sigma\mathbf D_{\mathbf S})]=\mathrm{tr}(\mathbf D^2_{\boldsymbol \Sigma})$.

\end{enumerate}
\section{Consistency of $Y_{1N}$, $Y_{2N}$ and $Y_{3N}$}\label{Appendix2}
Note that $\mathrm{E}[Y_{1N}]=\mathrm{E}[\mathrm{tr}(\mathbf S)]=\mathrm{tr}(\boldsymbol \Sigma)$. Under assumption~(\ref{CovMatAss}), derivations in \cite{Chen2010a} imply that
$$\frac{\mathrm{Var}[Y_{1N}]}{\mathrm{tr}^2(\boldsymbol \Sigma)}\leq \left[\frac{2+\max\{0,B\}}{N}+\frac{2}{N(N-1)}\right]\frac{\mathrm{tr}(\boldsymbol \Sigma^2)}{\mathrm{tr}^2(\boldsymbol \Sigma)} \leq \frac{3+\max\{0,B\}}{N} \rightarrow 0.$$
Thus $Y_{1N}$ is a ratio-consistent estimator to $\mathrm{tr}(\boldsymbol \Sigma)$. Similarly, it can be shown that $Y_{2N}$ and $Y_{3N}$ are unbiased estimators to $\mathrm{tr}(\boldsymbol \Sigma^2)$ and $\mathrm{tr}(\mathbf D^2_{\boldsymbol \Sigma})$ respectively, and that $\mathrm{Var}[Y_{2N}]/\mathrm{tr}^2(\boldsymbol \Sigma^2) \rightarrow 0$ and $\mathrm{Var}[Y_{3N}]/\mathrm{tr}^2(\mathbf D^2_{\boldsymbol \Sigma}) \rightarrow 0$ as $N \rightarrow \infty$. Hence, $Y_{2N}$ and $Y_{3N}$ are ratio-consistent estimators of  $\mathrm{tr}(\boldsymbol \Sigma^2)$ and $\mathrm{tr}(\mathbf D^2_{\boldsymbol \Sigma})$. 
\section{Alternative Formulas for $Y_{1N}$, $Y_{2N}$ and $Y_{3N}$}\label{Appendix3}
Note that
\begin{align*}
Y_{1N}&=U_{1N}-U_{4N}=\frac{1}{N}\sum_{i=1}^N \mathbf X^{T}_{i} \mathbf X_{i} -\frac{1}{P^N_2}\sum_{i \neq j}^{\ast} \mathbf X^{T}_{j}\mathbf X_{i}=\mathrm{tr}(\mathbf S).
\end{align*}
\cite{Himenoa2012} showed that
\begin{align*}
Y_{2N}&=U_{2N}-2U_{5N}+U_{6N}\\
      &=\frac{1}{P^N_2}\sum_{i \neq j}^{\ast} (\mathbf X^{T}_{i}\mathbf X_{j})^2-2\frac{1}{P^N_3}\sum_{i \neq j \neq k}^{\ast} \mathbf X^{T}_{i}\mathbf X_{j}\mathbf X^{T}_{i}\mathbf X_{k}\\
			&+\frac{1}{P^N_4} \sum_{i\neq j \neq k \neq l}^{\ast} \mathbf X_{i}\mathbf X^{T}_{j}\mathbf X_{k}\mathbf X^{T}_{l}\\
			&= \frac{N-1}{N(N-2)(N-3)}\left[(N-1)(N-2)\mathrm{tr}(\mathbf S^2)+\mathrm{tr}^2(\mathbf S)-NQ\right]
\end{align*}
where $$Q=\frac{1}{N-1}\sum_{i=1}^N \left[(\mathbf{X}_i-\bar{\mathbf{X}})^T(\mathbf{X}_i-\bar{\mathbf{X}})\right]^2.$$
Also it can be shown that 
\begin{align*}
Y_{3N}&=U_{3N}-2U_{7N}+U_{8N} \nonumber \\
      &=\frac{1}{P^N_2}\sum_{i \neq j}^{\ast} \mathrm{tr}(\mathbf X_{i}\mathbf X^T_{i} \circ \mathbf X_{j}\mathbf X^T_{j})-2\frac{1}{P^N_3}\sum_{i \neq j\neq k}^{\ast} \mathrm{tr}(\mathbf X_{i}\mathbf X^T_{i} \circ \mathbf X_{j}\mathbf X^T_{k})\\
			&+\frac{1}{P^N_4} \sum_{i\neq j \neq k \neq l}^{\ast} \mathrm{tr}(\mathbf X_{i}\mathbf X^{T}_{j} \circ \mathbf X_{k}\mathbf X^{T}_{l})\\
			&=\frac{1}{P^N_2}\sum_{a=1}^{p}\sum_{i\neq j}^{\ast} X^2_{ia}X^2_{ja}\\
			&-4\frac{1}{P^N_3}\left\{\sum_{a=1}^p\left(\sum_{i=1}^N X^2_{ia}\right)\left(\sum_{i=1}^{N-1}\sum_{j=i+1}^{N} X_{ia}X_{ja}\right)-\sum_{a=1}^{p}\sum_{i\neq j}^{\ast} X^3_{ia}X_{ja}\right\} \\
			&+\frac{2}{P^N_4}\left\{2\sum_{a=1}^p\left(\sum_{i=1}^{N-1}\sum_{j=i+1}^{N} X_{ia}X_{ja}\right)^2-\sum_{a=1}^{p}\sum_{i\neq j}^{\ast} X^2_{ia}X^2_{ja}  \right\}.
\end{align*}
Together these results reduce the computational cost for $Y_{1N}$, $Y_{2N}$ and $Y_{3N}$ from $O(N^4)$ to $O(N^2)$.

\bibliographystyle{plainnat}
\bibliography{dissrefer.bib}

\end{document}